\documentclass[aps,prd,floatfix,nofootinbib,showpacs,twocolumn,10pt]{revtex4-1}

\usepackage{amssymb}
\usepackage[intlimits]{amsmath}
\usepackage{amsfonts}
\usepackage{dsfont}
\usepackage{subfigure}
\usepackage[usenames,dvipsnames]{color}
\usepackage{graphicx}
\usepackage{natbib}
\usepackage{multirow}
\usepackage{bbm}

\usepackage{slashed}

\usepackage[usenames,dvipsnames,svgnames,table]{xcolor}

\newcommand{\Tr}{\ensuremath{\operatorname{Tr}}}

\newcommand{\be}{\begin{equation}}
\newcommand{\ee}{\end{equation}}
\newcommand{\ba}{\begin{eqnarray}}
\newcommand{\ea}{\end{eqnarray}}
\newcommand{\bi}{\begin{itemize}}
\newcommand{\ei}{\end{itemize}}

\def\cA{{\cal A}}

\def\cD{{\cal D}}

\def\cF{{\cal F}}

\def\cH{{\cal H}}

\def\cO{{\cal O}}

\newcommand{\p}{\partial}

\begin{document}
\title{Spectral dimensions from the spectral action}

\author {Nat\'alia Alkofer}
\email{N.Alkofer@science.ru.nl}
\author {Frank Saueressig}
\email{F.Saueressig@science.ru.nl}

\affiliation{ Institute for Mathematics, 
Astrophysics, and Particle Physics (IMAPP), Radboud University Nijmegen, \\
Heyendaalseweg 135, 6525 AJ Nijmegen, The Netherlands}

\author{Omar Zanusso}
\email{omar.zanusso@uni-jena.de}

\affiliation{Theoretisch-Physikalisches Institut, Friedrich-Schiller-Universit\"{a}t Jena,
Max-Wien-Platz 1, 07743 Jena, Germany}

\date{\today}


\begin{abstract}
The generalised spectral dimension $D_{ S}(T)$ provides a powerful tool for comparing 
different approaches to quantum gravity. In this work, we apply this formalism to the classical 
spectral actions obtained within the framework of almost-commutative geometry. Analysing the 
propagation of spin-0, spin-1 and spin-2 fields, we show that a non-trivial spectral dimension 
arises already at the classical level. The effective field theory interpretation
of the spectral action yields plateau-structures interpolating between a fixed spin-independent $D_{ S}(T) = d_S$ for short and 
$D_{ S}(T) = 4$ for long diffusion times $T$. Going beyond effective field theory
 the spectral dimension 
is completely dominated by the high-momentum properties of the spectral action, yielding 
$D_{ S}(T)=0$ for all spins.
Our results support earlier claims that high-energy bosons do not 
propagate.
\end{abstract}

\maketitle
\section{Introduction}
\label{sec:intro}
The spectral action principle \cite{Chamseddine:1991qh,Chamseddine:1996zu} provides
a framework for unifying gravity and elementary particle physics on the basis  of non-commutative
geometry \cite{Connes:1994yd}, see
\cite{connesreview,Jureit:2007qm,Sakellariadou:2013ve,vandenDungen:2012ky} for reviews. A key
ingredient in the construction is the spectral action
\cite{Chamseddine:1991qh,Chamseddine:1996zu}
\begin{equation}
S_{\chi,\Lambda}=\Tr \bigl[ \chi(\cD^2/\Lambda^2) \bigr] 
\label{SpAct1}
\end{equation}
where $\chi$ is a positive function, $\cD$ is a Dirac operator on 
a non-commutative geometry,
$\Lambda$ is a suitable cutoff scale, and the trace indicates the sum over eigenvalues
of $\cD$. 
Amongst non-commutative geometries almost-commutative ones lead, 
for suitable choices of the almost-commutative manifold, to a spectral action which can give
rise to the standard model of particle physics
minimally coupled to
gravity
\cite{Connes:2006qv,Chamseddine:2006ep,Chamseddine:2007ia,Chamseddine:2012sw,Stephan:2013rna}.
These models may also be extended to include physics beyond the standard model
\cite{Chamseddine:2010ud,Devastato:2013oqa,Chamseddine:2013sia,Chamseddine:2013rta} or
supersymmetry \cite{Ishihara:2013asa,Beenakker:2014yla,Beenakker:2014zla,Beenakker:2014ama}.
Their renormalization has been studied in
\cite{vanSuijlekom:2011uu,vanSuijlekom:2011kc,vanSuijlekom:2012xb,Estrada:2012te,Suijlekom:2014ata}
and the phenomenological implications of the resulting effective actions have been carried out,
e.g., in \cite{Nelson:2010ru,Lambiase:2013dai,Chamseddine:2014nxa}.
A more detailed discussion of
the cutoff $\Lambda$ may be found in \cite{D'Andrea:2013nda}, and
the generalization to non-commutative spaces built from non-associative algebras has been pursued
in \cite{Farnsworth:2013nza}.    

In this work, we follow up on Refs.\
\cite{Kurkov:2013kfa,Iochum:2011yq} and study the properties of the spectral action beyond the framework of
effective field theory. More specifically, we construct the generalized spectral dimension
$D_S(T)$ \cite{Ambjorn:2005db,Sotiriou:2011mu,Reuter:2011ah} 
resulting from the spectral action principle and compare our findings with other
approaches to quantum gravity.

The basic idea underlying the concept of the generalized spectral dimension $D_S(T)$ is that a test
particle diffusing on a given background probes certain features of the background. 
 By now, this ``observable'' has been computed in
many approaches to quantum gravity and a number of quantum gravity inspired models. 
A common feature shared by many of these studies is 
 a ``dynamical dimensional reduction'' from a classical spacetime
with $D_S(T) = 4$ at macroscopic  scales to $D_S(T) = 2$ \cite{Carlip:2009kf,Carlip:2012md}. The widespread
use of $D_S(T)$ throughout the quantum gravity community then calls for a detailed understanding
which (quantum) features of a model are actually encoded in $D_S(T)$.

In the literature, there are  essentially three approaches to compute
$D_S(T)$. For the piecewise linear geometries approximating spacetime within Monte Carlo simulations
of the gravitational  partition sum, one sets up a random walk on the effective quantum geometry. This
allows a direct measurement of $D_S(T)$ from the return probability of the random walker, thereby
characterizing the fractal features of the geometry. A prototypical example for this setup is provided
by the Causal Dynamical Triangulations (CDT) program \cite{Ambjorn:2005db,Benedetti:2009ge}. The
second way towards obtaining  a non-trivial $D_S(T)$ starts from a theory where the propagator of the
test particle is already modified at the \emph{classical} level. It is this feature that gives rise to
the non-trivial spectral dimension in Ho\v{r}ava-Lifshitz gravity
\cite{Horava:2009if,Sotiriou:2011mu,Sotiriou:2011aa}, general Lorentz-violating theories
\cite{Amelino-Camelia:2013gna,Amelino-Camelia:2013cfa}, or fractional quantum field theory
\cite{Calcagni:2010pa,Calcagni:2013sca}. Finally, $D_S(T)$ may be modified through non-trivial quantum
fluctuations of spacetime predicted by a fundamental theory of gravity. This set includes the
multifractal spacetimes occurring in the gravitational Asymptotic Safety program
\cite{Lauscher:2005qz,Reuter:2011ah,Reuter:2012xf,Rechenberger:2012pm,Calcagni:2013vsa}, the
microscopic structure of spacetime within Loop Quantum Gravity
\cite{Modesto:2008jz,Modesto:2009qc,Magliaro:2009if,Caravelli:2009gk,Modesto:2009kq,Calcagni:2013dna}, 
Causal Set Theory \cite{Eichhorn:2013ova} {
 or the propagation of particles on
$\kappa$-Minkowski
\cite{Benedetti:2008gu,Arzano:2014jfa}
and other non-commutative spaces \cite{Alesci:2011cg}.
}

For the classical spectral action, the non-trivial spectral dimension originates from the non-canonical momentum dependence of the theory's propagators.
 The essentially new feature, making
this model worthwhile to study, is that the (inverse) full propagators of the model are non-analytic
functions of the momentum. This will lead to a rather surprising behaviour of the spectral dimension
computed in this framework.


The rest of this work is devoted towards studying these novel features. In Sects. \ref{sec:SpecAct1} and \ref{sec:specdim} we
review the relevant aspects of the spectral action and the generalized spectral dimension, respectively. The
spectral dimension resulting from the spectral action is computed in Sect. \ref{sec:IV} and we
conclude with a brief summary and outlook in Sect. \ref{sec:Concl}.

\section{The spectral action and \\ its bosonic propagators}
\label{sec:SpecAct1} 
We start by reviewing the spectral action and its connection to
the heat-kernel mainly following \cite{vandenDungen:2012ky,Kurkov:2013kfa}. 
The basic ingredient for constructing the spectral action is an almost-commutative
manifold $M \times F$. Here $M$ is a Riemannian spin manifold which plays the role of
the Euclidean spacetime and $F$ a finite, generally non-commutative, space encoding
the internal degrees of freedom. The geometry of $M$ has a 
operator 
algebraic description in terms of the canonical triple 
\be
M := (C^\infty(M), L^2(M,S), \cD)
\ee
where $C^\infty(M)$ is the set of smooth functions on $M$, $L^2(M,S)$ is the Hilbert space
of square-integrable spinors on $M$, and $\cD$ is the (Euclidean) Dirac operator
acting on this Hilbert space. Similarly, the geometry of $F$ can be captured by a triple
\be
F := \left( \cA_F, \cH_F, D_F \right) \, .  
\ee
where $\cH_F$ is a finite-dimensional Hilbert space of complex dimension $N$, $\cA_F$ is an algebra 
of $N \times N$ matrices acting on $\cH_F$, and $D_F$ is a hermitian
operator, given by a hermitian $N\times N$ matrix.

In order to illustrate the propagation of particles resulting from this setup, it suffices to chose 
the simplest internal space, taking $F$ as a single point. In this case the triple 
$F = ({\mathbb C}, {\mathbb C}, 0)$ and
the Dirac operator on the product space reduces to the one on $M$. Concretely, we will consider
\be
\cD = \slashed{D} + \gamma_5 \, \phi \, , 
\ee
where the covariant derivative
\be
\slashed{D} = i \gamma^\mu \, \left( \nabla_\mu ^{LC} + i A_\mu \right) \, ,
\ee
contains the Levi-Cevita spin connection and the gauge potential $A_\mu$.
Consequently, the resulting
spectral action comprises a spin-2 field, the graviton, 
a massless $U(1)$ gauge field $A_\mu$ with field strength $F_{\mu\nu} = \p_\mu A_\nu - \p_\nu A_\mu$ 
and a scalar $\phi$. In the sequel, we will study   the transverse traceless fluctuations 
$h_{\mu\nu}$ with $\p^\mu h_{\mu\nu} = 0$, $\delta^{\mu\nu} h_{\mu\nu} = 0$ only
and impose the Landau gauge for the spin-1 field $\p^\mu A_\mu = 0$,
limiting ourselves to the propagation of the physical degrees of freedom.

The operator $\cD^2$ appearing in the spectral action (\ref{SpAct1}) can then be cast
into the standard form of a Laplace-type operator
\begin{equation} 
\cD^2 = -(\nabla^2 +E) 
\end{equation}
with the endomorphism $E$ given by
\begin{equation}
E= -i \gamma^\mu \gamma_5 \nabla_\mu \phi - \phi^2 - \frac 1 4 R + \frac i 4
[\gamma^\mu, \gamma^\nu] F_{\mu\nu} .
\end{equation}
Moreover, the curvature of $\nabla_\mu=\nabla_\mu^{LC}+iA_\mu$ is given by
\begin{equation}
\Omega_{\mu\nu}  := [\nabla_\mu , \nabla_\nu ] = - \frac 1 4 \gamma^\rho \gamma^\sigma
R_{\rho\sigma\mu\nu} +i F_{\mu\nu} .
\end{equation} 

At this stage, it is illustrative to follow \cite{Kurkov:2013kfa} and consider the special case
where the generating function in eq.\ \eqref{SpAct1} is given by $\chi(z) = e^{-z}$
In this case, 
the spectral action (\ref{SpAct1}) coincides with the heat trace
\begin{equation}
S_{\chi,\Lambda} = \Tr \left( e^{-{ t} \cD^2}\right) = K(\cD^2, { t} )
\quad \mathrm{with} \quad { t} := \Lambda^{-2} \, , 
\label{SpAct}
\end{equation}
which is a well-studied object, see, e.g., 
\cite{Vassilevich:2003xt,Barvinsky:1987uw,Barvinsky:1990up,Iochum:2011yq,Codello:2012kq}.

\begin{figure}
 \includegraphics[width=\columnwidth]{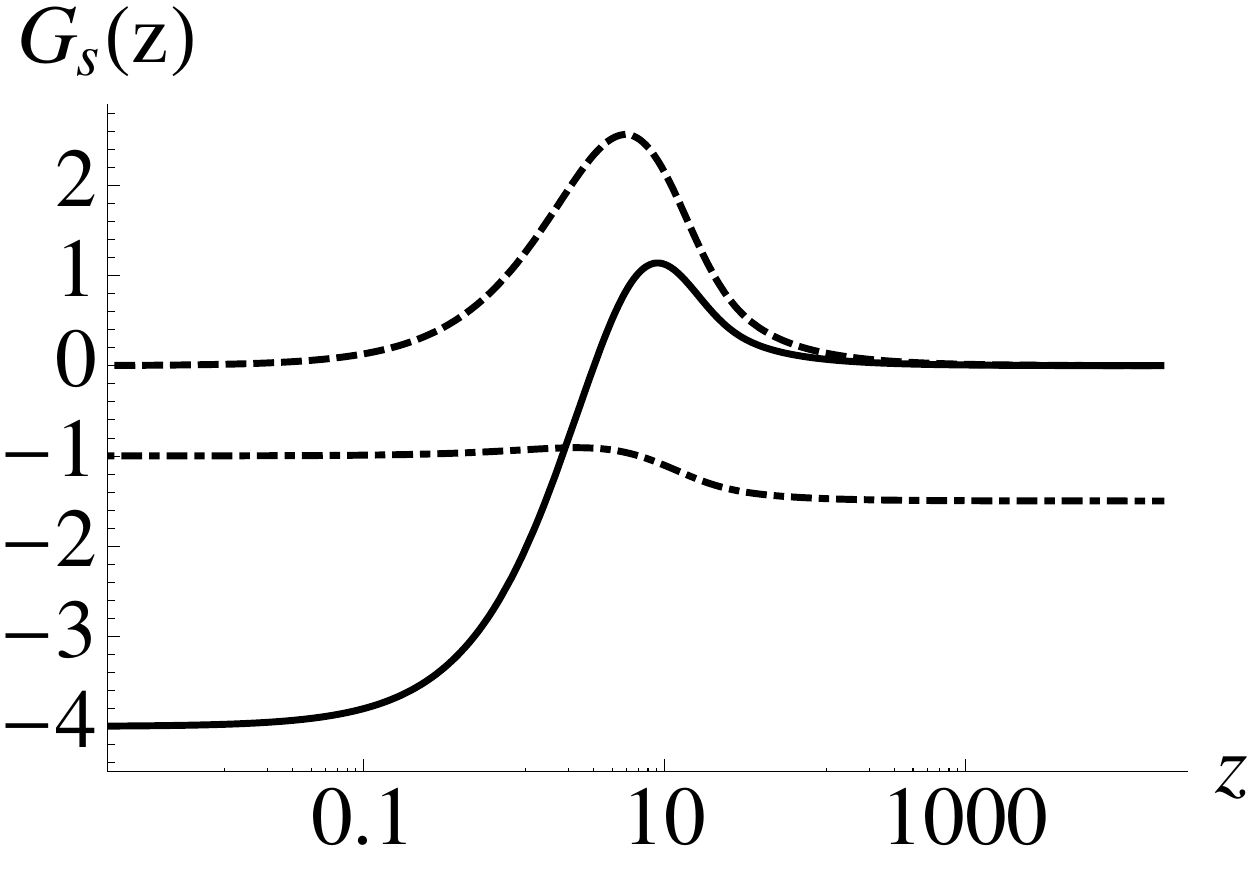}
 \caption{\label{figsd_F012_F} Illustration of the momentum dependence of the structure 
functions ${ G}_{s}(z)$ (\ref{Gsfct}): 
spin 0 - solid thick line, spin 1- dashed thick line,  and spin 2 - dash-dotted
thick line. } 
 \end{figure}

As we are interested in the propagation of fields we will extract the (inverse)
 propagators of the matter fields by expanding \eqref{SpAct} up to second order
in $\phi$, $A_\mu$. The inverse propagator in the gravity sector is obtained 
by expanding $g_{\mu\nu}$ around flat (Euclidean) space
\be\label{expmet}
g_{\mu\nu}= \delta_{\mu\nu} + \, \Lambda^{-1} \, h_{\mu\nu} \,   . 
\ee
The inclusion of $\Lambda$ ensures that $h_{\mu\nu}$ has the same mass-dimension
as the matter fields and gives rise to the canonical form of the graviton propagator.
Comparing \eqref{expmet} with the stand expansion of $g_{\mu\nu}$ used in perturbation theory,
$g_{\mu\nu}= \delta_{\mu\nu} + \sqrt{16 \pi G_{\rm N}} \,  h_{\mu\nu} $ with $G_{\rm N}$ being 
Newton's constant,
identifies the natural scale for $\Lambda$ as the Planck mass $m_{\rm Pl} = (8 \pi G_{\rm N})^{-1/2}$
(also see \cite{Devastato:2013wza} for a related discussion). 
A lengthy but in principle straightforward calculation
\cite{Kurkov:2013kfa,Barvinsky:1987uw,Barvinsky:1990up,Iochum:2011yq,Codello:2012kq}
then yields the expression for the inverse propagators of the physical fields including the full-momentum dependence
\begin{eqnarray}
\label{Kmod}
&& K^{(2)}(\cD^2,{ t})  =   \int d^4x \Biggl[ 
 \phi F_{0} (- { t} \partial^2) \phi \nonumber \\ 
 && \quad \; +  \,  A_\mu F_1(- { t} \partial^2)  A_\mu 
+ \Lambda^{-2} \, h_{\mu\nu} F_{2} (- { t} \partial^2) h_{\mu\nu}
\Biggr] .
\label{K2}
\end{eqnarray}
Here $K^{(2)}$ indicates that we retained the second order of the fields only and
we explicitly exhibit the factor $\Lambda^{-2}$ coming from the expansion \eqref{expmet}. The
structure functions $F_{s}$ coincide with the standard heat kernel result for spin-$s$ fields and 
are given by
\begin{eqnarray}
F_{0} (z) &=& \frac { { t}^{-1}} {(4\pi)^2} \left( - 4 +  2 z h(z) \right),
\label{F0} \\
F_{1} (z) &=&  \frac { { t}^{-1}} {(4\pi)^2} \left(  -4 + 4 h(z) + 2 z h(z) \right),
\label{F1} \\
F_{2} (z) &=& \frac { { t}^{-2}} {(4\pi)^2} \left(  - 2  + h(z) +  \frac 1 4 zh(z)\right),
\label{F2}
\end{eqnarray}
with
\begin{equation}
h(z)= \int_0^1 d\alpha \,e^{-\alpha (1-\alpha) z} .
\end{equation}
Note that these functions are non-analytic in 
\be
z= \frac{p^2}{\Lambda^2} = { t} \, p^2.
\ee
The inverse of these structure functions provide the classical propagators of the theory.
For illustration, we show the ``reduced'' structure functions
\be\label{Gsfct}
G_s(z) = (4 \pi)^2 \, { t}^{\alpha_s} \, F_s(z)
\ee
with $F_s(z)$ defined in eqs.\ (\ref{F0}) - (\ref{F2}) and $\alpha_s = (1,1,2)$ in Fig.\ \ref{figsd_F012_F}. 

 \begin{figure}
 \includegraphics[width=\columnwidth]{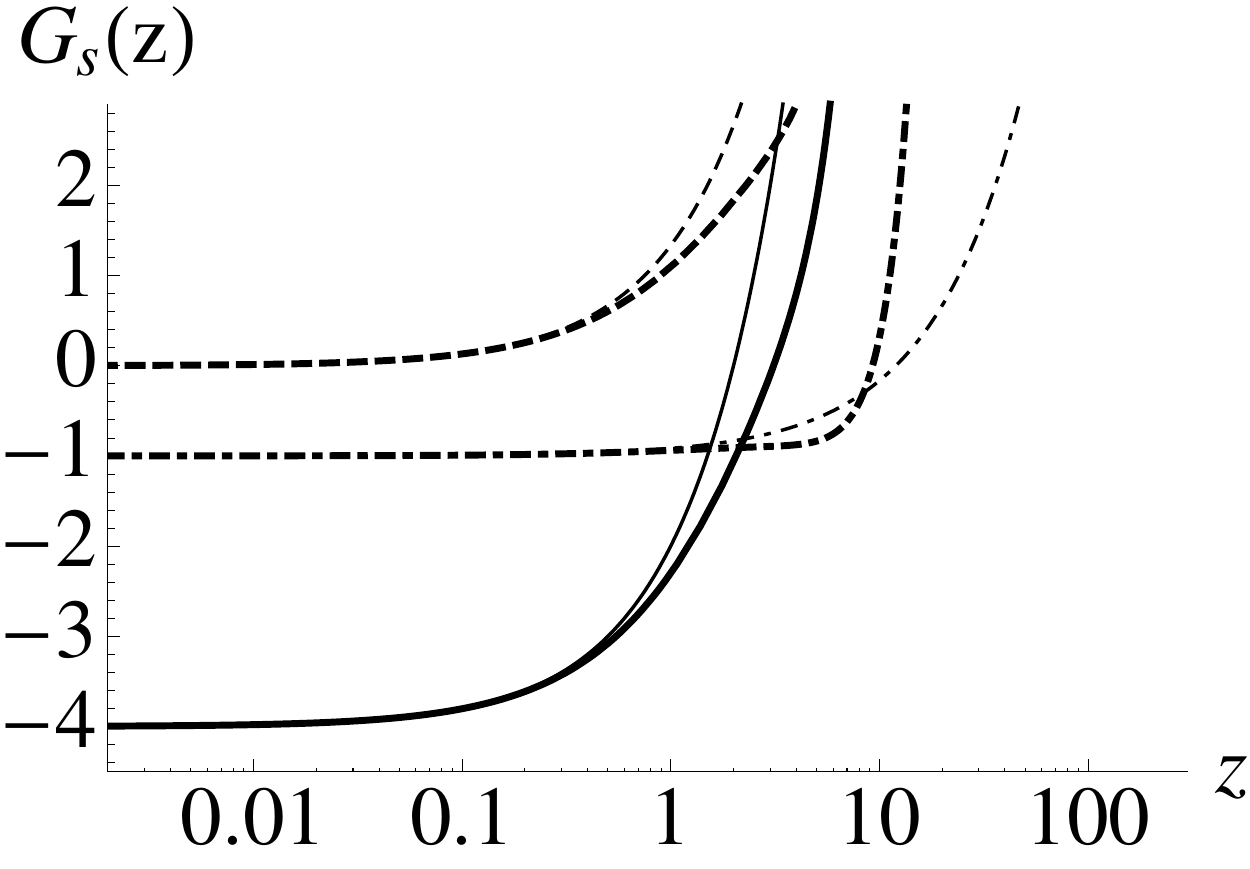}
 \caption{\label{proptrunc} Propagators obtained from truncating $G_s(z)$ at $z^3$, 
 following the spirit of effective field theory (spin 0 - solid thick line, spin 1- dashed 
 thick line,  and spin 2 - dash-dotted). The expansions up to linear order in $z$ are shown as 
 corresponding 
thin lines.} 
 \end{figure}
 
 The structure functions (\ref{F0}) - (\ref{F2}) posses an early-time expansion for 
 momenta $p^2 \ll \Lambda^2$, i.e., $z < 1$. Using that
\be
h(z) = 1- \frac z 6+ \frac{z^2}{60}-\frac{z^3}{840} + \cO(z^4) 
\ee
one finds
\be\label{earlytime}
\begin{split}
(4\pi)^2 \, { t} \, F_{0} (z) =& - 4 + 2 z  - \frac{z^2}{3}  +
\frac{z^3}{30} + \ldots \, ,\\
(4\pi)^2 \, { t} \,  F_{1} (z) =& \frac 4 3 z -  \frac {4 }{15} z^2 +
\frac {1}{35} z^3  + \ldots \,  ,\\
(4\pi { t} )^2  \,  F_{2} (z) =& - 1 + \frac z{12} -  \frac {z^2}{40} +
\frac {z^3}{336} 
+ \ldots   \, .
\end{split}
\ee
The early time expansion of the structure functions (truncated at $\cO(z^3)$ and $\cO(z)$, 
respectively) is
shown in Fig.\ \ref{proptrunc}. Comparing the result to the functions $F_s(z)$ including the 
full $z$-dependence
it is easily seen that the truncation drastically modifies the behaviour of the (inverse) 
propagators
for large momenta. While 
the truncated $F_s(z)$ diverge the full structure functions remain finite. As we will see in 
Sect.\ \ref{sec:IV}, this feature will have a drastic effect
on the spectral dimension of the theory.

At this stage, the following remarks are in order. The constant term appearing
in the expansion of $F_{0}$ plays the role of a mass term for $\phi$. The sign thereby indicates 
that the squared mass is negative. This is a remnant of the fact that the scalar $\phi$ acquires a 
non-trivial vacuum expectation value via the Higgs mechanism 
\cite{Chamseddine:1991qh,Chamseddine:2012sw,Estrada:2012te}. Since $K^{(2)}$, by construction, 
contains the terms quadratic in $\phi$ only
the scalar potential is not included in \eqref{K2} so that the stabilisation of $\phi$ 
cannot be demonstrated in this approximation. The constant
term in $F_2$ indicates the presence of a positive cosmological constant, acting like a 
mass-term for the graviton, while the structure of $F_1$ indicates that the gauge field 
remains massless.

In order to understand the behaviour of the theory at high energies, it is also useful to 
carry out the late-time expansion of the structure functions, capturing the behaviour 
for $z \gg 1$. In this case
\be
h(z) = \frac 2 z+ \frac 4{z^2}+ \frac{24}{z^3}+ \frac{240}{z^4}+ \ldots
\ee
which yields
\be\label{latetime}
\begin{split}
(4\pi)^2 \, { t} \,  F_{0} (z) =&  \frac 8 z + \frac {48}{z^2} + \frac{480}{z^3}+\ldots \, ,\\
(4\pi)^2 \, { t} \,  F_{1} (z) =& \frac {16} {z} + \frac{64}{z^2} + \frac {576}{z^3} +\ldots \, ,\\
(4\pi { t} )^2  \,  F_{2} (z) =&  - \frac 3 2 + \frac 3 z +  \frac {10}{z^2} + \frac {84}{z^3}  +
\ldots \, . \\
\end{split}
\ee

A typical viewpoint adopted in the spectral action approach to particle physics consider
the actions generated by \eqref{SpAct1} as effective actions which should be truncated
at a certain power of the cutoff $\Lambda^{-2}$. In this case the early-time expansion
\eqref{earlytime} allows to construct the effective action resulting from
an arbitrary function $\chi$. This uses the fact that the Trace \eqref{SpAct1} can
be related to the heat kernel \eqref{SpAct} using
\be\label{mellin1}
\begin{split}
S_{\chi, \Lambda} = & \, \Tr \bigl[ \chi (\cD^2/\Lambda^2 ) \bigr] \\
= & \int_0^\infty dy \, \tilde{\chi}(y) \, \Tr \bigl[ e^{-y \cD^2/\Lambda^2} \bigr] \, ,
\end{split}
\ee 
where $\tilde{\chi}(y)$ is the inverse Laplace-transform of $\chi(z)$. 
Evaluating the operator trace in \eqref{mellin1} based on the early-time expansion then yields
the systematic expansion of $S_{\chi, \Lambda}$ in (inverse) powers of the cutoff. 
The $\chi$-dependent coefficients
in this expansion are given by
\be\label{Qdef}
\begin{split}
Q_n[\chi] \equiv & \, \int_0^\infty dy \, y^{-n} \, \tilde{\chi}(y)  
\end{split}
\ee
and can be computed by standard Mellin-transform techniques \cite{Codello:2008vh}.
Thus the $Q_n \equiv Q_n[\chi]$ are real numbers which are normalized such that
 $Q_n=1$ for $\chi=\exp(- { t} z)$.
 
The part of the spectral action containing the terms quadratic in the fields
is then given by
\begin{equation}
\begin{split}
S^{(2)}_{\chi,\Lambda} = \frac{\Lambda^2}{(4\pi)^2} \int d^4x &\Bigl[  
  \phi \, \cF_{0,\chi}\left(-\partial^2/\Lambda^2\right) \phi \\
 & +  A_{\mu} \cF_{1,\chi}\left(-\partial^2/\Lambda^2\right)  A_{\mu} \label{Kmod2} \\
 & + h_{\mu\nu} \cF_{2,\chi}\left(-\partial^2/\Lambda^2\right) h_{\mu\nu} 
\Bigr] \, . 
\end{split}
\end{equation}
with
\be\label{earlytime2}
\begin{split}
\cF_{0,\chi} (z) =& - 4 Q_1 +2 {Q_0} z - \frac{Q_{-1}}{3}  z^2  + \frac{Q_{-2}}{30} z^3 + \dots \, ,\\
\cF_{1,\chi} (z) =& \frac {4Q_0} 3 z -  \frac {4 Q_{-1}}{15} z^2 + \frac {Q_{-2}}{35} z^3  + \dots \,  ,\\
\cF_{2,\chi} (z) =& - Q_2 + \frac {Q_1}{12}z -  \frac {Q_{0}}{40}z^2 + \frac {Q_{-1}}{336}z^3 + \dots \, .\\
\end{split}
\ee
The momenta \eqref{Qdef} can be adjusted by choosing a suitable function $\chi$.
Note, however, that the $Q_n$'s appearing in the matter and gravitational sector of 
\eqref{Kmod2} \emph{cannot be adjusted} independently, however, since they are
generated by the same function $\chi$. 

We close this section by discussing possible truncations of the expansion \eqref{Kmod}: 

\noindent
{\it Truncating the moments $Q_n$.}
From the mathematical viewpoint it is tempting to choose a generating function $\chi$ whose moments
$Q_n[\chi]$ vanish for all values $n \ge n_{\rm max}$. 
This leads to the rather peculiar property that the highest powers of  $-\p^2$ appearing in the 
matter and gravitational sector 
\emph{come with opposite signs}. In other words
adjusting the $Q_n$ in such a way that the propagators in the matter sector are stable at high 
momenta implies an instability
of the gravitational propagator and vice versa.
Thus ``truncating'' the theory by adjusting the momenta $Q_n$  
gives rise to a dynamical 
instability of the
theory.

\noindent
{\it The effective field theory viewpoint.}
A similar (though not equivalent) strategy interprets the expansion \eqref{Kmod}
as an effective field theory, which should be truncated at a given power of the cutoff $\Lambda$. 
Retaining the relevant and marginal operators then provides a good description
of the physics as long as $-p^2/\Lambda^2 \ll 1$. While it is possible
to systematically compute quantum corrections to an effective action, this expansion breaks down
if the momenta are of the order of the Planck scale. A detailed analysis then reveals
that the $Q_n$'s can be adjusted in such a way that all propagators of the theory
are stable. Thus we will focus on this case in the sequel.
 
\section{The generalized spectral dimension}
\label{sec:specdim} 
The motivation for studying the (generalized) spectral dimension
associated with a particle physics model comes from the idea
that a test particle diffusing on a given fixed background
feels certain features of this background as, e.g., its dimension.
For a spin-less test-particle performing a Brownian random 
walk on a Riemannian manifold with metric $g_{\mu\nu}$, the
diffusion process is described by the heat kernel $K_g(x,x^\prime;T)$
which gives the probability density for a particle diffusing from the 
point $x$ to $x^\prime$ in the diffusion time $T$. The heat kernel
satisfies the heat equation
\be\label{diffeq}
\begin{split}
& \left( \p_T + \Delta_g \right) K_g(x,x^\prime;T) = 0 \, , \\
& K_g(x,x^\prime;0) = \delta(x-x^\prime) \, . 
\end{split}
\ee
where $\Delta_g \equiv -D^2$ is the Laplace-Beltrami operator. In flat space, the solution
of this equation is
\be\label{flatspace1}
K(x,x^\prime;T) = \int \frac{d^dp}{(2\pi)^d} \, e^{ip \cdot(x-x^\prime)} \, e^{-p^2 T} \, . 
\ee
In general $K_g(x,x^\prime;T)$ is the matrix element of the operator $\exp(-T \Delta_g)$.
For the diffusion process, its trace per volume gives the averaged return probability
\be
\begin{split}
{\cal P}_g(T) = & \, V^{-1} \, \int d^dx \sqrt{g(x)} K_g(x,x;T) \\
= & \, V^{-1} \, {\rm Tr} \, \exp(-T \Delta_g) \, , 
\end{split}
\ee
measuring the probability that the particle returns to its origin after a diffusion time $T$.
Here $V \equiv \int d^dx \sqrt{g(x)}$  denotes the total volume. For the flat-space solution
\eqref{flatspace1}
\be
{\cal P}(T) = ( 4 \pi \, T)^{-d/2} \, . 
\ee

The (standard) spectral dimension $d_S$ is defined as the $T$-independent
logarithmic derivative
\be\label{spec1}
d_S \equiv -2 \lim _{T\to 0} \, \frac{ \partial \ln {\cal P}(T)}{\partial \ln T} \, . 
\ee
On smooth manifolds $d_S$ agrees with the topological dimension of the manifold $d$.
In order to also capture the case of diffusion processes exhibiting multiple scaling 
regimes, it is natural to generalize the definition \eqref{spec1} to the $T$-dependent
spectral dimension
\be\label{spec2}
D_S(T) \equiv -2 \, \frac{ \partial \ln {\cal P}(T)}{\partial \ln T} \, . 
\ee
 
In the classical spectral action \eqref{SpAct1} the propagation of the test
particles on a flat Euclidean background is modified by the higher-derivative
terms entering into the (inverse) propagators of the fields. In eq.\ \eqref{diffeq}
this effect can readily be incorporated by replacing the Laplace-Beltrami operator
by the inverse propagators
\be\label{diffeq2}
\begin{split}
& \left( \p_T + F(-\p^2) \right) K_g(x,x^\prime;T) = 0 \, , \\
& K_g(x,x^\prime;0) = \delta(x-x^\prime) \, .
\end{split} 
\ee
The solution of this equation can again be given in terms of its Fourier transform
\be\label{flatspace2}
K(x,x^\prime;T) = \int \frac{d^dp}{(2\pi)^d} \, e^{ip \cdot(x-x^\prime)} \, e^{- F(p^2) T} \, . 
\ee
Notably, given a generic function $F(p^2)$ there is no guarantee that the resulting
heat-kernel is positive semi-definite thereby admitting an interpretation as probability
density. This ``negative probability problem'' has been discussed in detail \cite{Calcagni:2013vsa,Calcagni:2014wba},
concluding that the spectral dimension remains meaningful.  

The ${\cal P}(T)$ resulting from \eqref{flatspace2} is given by
\be\label{returnprop}
{\cal P}(T) = \int \frac{d^dp}{(2\pi)^d} \, e^{- F(p^2) T} \, . 
\ee
and may still admit the interpretation of a (positive-semidefinite) return probability
even in the case where a probability interpretation of $K(x,x^\prime;T)$ fails. The generalized
spectral dimension may then be obtained by substituting the inverse propagators from eq.\ \eqref{Kmod} and evaluating
\eqref{spec2} for the corresponding return probabilities.
 
Following the ideas of \cite{Amelino-Camelia:2013gna,Amelino-Camelia:2013cfa} the spectral dimension
arising from \eqref{returnprop} permits an interpretation as the Hausdorff-dimension of the
theory's momentum space. Provided that the change of coordinates $k^2 = F(p^2)$ is bijective, 
the inverse propagator in the exponential can be traded for a non-trivial measure on momentum space
\be
P(T) = \frac{{\rm Vol}_{S^d}}{(2\pi)^d} \, \int k dk \, \frac{\left(F^{-1}(k^2)\right)^{(d-2)/2}}{F^\prime(p^2)} \, e^{-T k^2} \, .  
\ee
Here we invoked the inverse function theorem where $F^\prime(p^2)$ is understood as the derivative of $F(z)$ with respect to its argument,
evaluated at $p^2 = F^{-1}(k^2)$. Thus eq.\ \eqref{returnprop} is equivalent to a particle with canonical inverse propagator, $F(p^2) \propto p^2$
in a momentum space with non-trivial measure. This picture also provides a meaningful 
interpretation of $D_S(T)$ even in the case where the model is purely classical so that the non-trivial spectral dimension
cannot originate from properties of an effective quantum spacetime.

Before embarking on the computation of the spectral dimensions
resulting from the spectral action it is illustrative to consider the case where the inverse momentum-space propagator $F(p^2)$ contains a mass term $m^2 = F(p^2)|_{p^2 = 0}$. In this case
it is useful to split off the  massless part from $F(p^2)$ and write  
\be\label{masssplit}
F(p^2) = F^{(0)}(p^2) + m^2 \, . 
\ee
Based on $F^{(0)}(k^2)$ we can then introduce 
 the return probability
\be\label{P0exp}
P^{(0)}(T) \propto \int \frac{d^4p}{(2\pi)^4 } \, e^{-TF^{(0)}(p^2)} 
\ee
together with the spectral dimension seen by the massless field
\be\label{Ds0}
D_S^{(0)}(T) \equiv -2  T \frac \partial {\partial T} \ln {\cal P}^{(0)}(T) \, .
\ee
Substituting \eqref{masssplit} into the return probability \eqref{returnprop}
 and extracting the mass-term from the integral
it is straightforward to establish
\be\label{d0def}
D_S(T) = 2 \, m^2 \, T + D_S^{(0)}(T) \, . 
\ee
Thus a mass-term just leads to a linear contribution in $D_S(T)$ and does not encode
non-trivial information on the propagation of the particle. Thus we limit ourselves
to the study of $D_S^{(0)}(T)$ in the sequel.

\section{The spectral dimension from the spectral action}
\label{sec:IV}
Based on the discussion of the last two sections, 
it is now straightforward to compute the 
spectral dimensions from the spin-dependent propagators provided by the
spectral action. We will start by investigating
the truncated propagators based on eqs.\ \eqref{Kmod} and \eqref{Kmod2} in Subsection \ref{sec:SpecAct2}
before including the full momentum dependence
in Subsection \ref{sect.IVb}. 

\subsection{Effective field theory framework}
\label{sec:SpecAct2}
In the effective field theory interpretation of \eqref{Kmod2} the functions
$\cF_{s,\chi}$ are truncated at a fixed power of the cutoff $\Lambda$. The resulting
massless parts of the bosonic propagators then become polynomials in
the particles momentum,
\be\label{fexp}
\cF_s^{(0)}(p^2) = \sum_{n=1}^{N_{\rm max}} \, a_n^s \, (p^2)^n \, , 
\ee
with obvious relations among the polynomial coefficients $a_n$ and the numbers $Q_n$
\eqref{earlytime2}. Limiting the expansion to the marginal and relevant operators,
coming with powers of the cutoff $\Lambda^{2n}$, $n \le 2$, fixes $N_{\rm max} = 1$ and
all propagators retain their standard $p^2$-form. Also taking into account power-counting
irrelevant terms containing inverse powers of the cutoff adds further 
powers to the polynomial \eqref{fexp}. Thus the propagators include higher powers of momentum
in this case.

A positive semi-definite spectral dimension $D_S^{(0)}$ requires a positive function
$\cF_s^{(0)}$. This requirement puts constraints on the signs of the momenta $Q_n$ appearing 
in eq.\ \eqref{earlytime2}. In particular $a_1^s > 0$ is required for obtaining classical propagators
at low energies while $a_{N_{\rm max}}^s > 0$ is needed for stability at high energies.

\begin{figure}
 \includegraphics[width=\columnwidth]{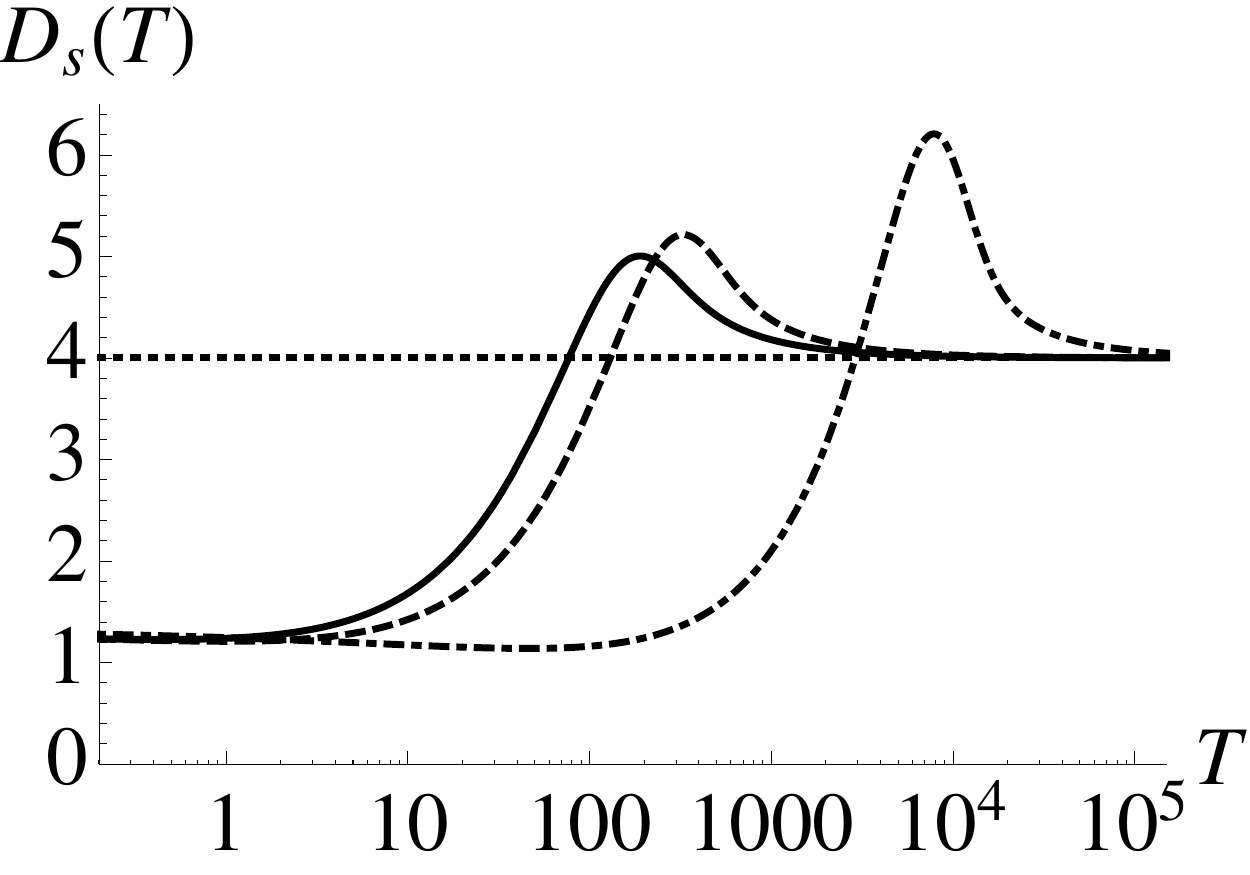}
\caption{\label{figsd_F012_D} The spin-dependent spectral dimension $D_S^{(0)}(T)$
obtained from \eqref{fexp} with $N_{\rm max} = 1$ (dotted line) and $N_{\rm max} = 3$,
$Q_n = 1$, $n=1,0,-1,-2$ for spin 0 - solid line, spin 1- dashed line,  and spin 2 - dashed-dotted
line. The cross-over scale is set by ${ t} = \Lambda^{-2}$ and has been normalized to $ { t} =1$.
}
\end{figure}

The asymptotic behavior of $D_S^{(0)}$ for short (long) diffusion time $T$ 
is governed by the highest (lowest) power of $p^2$ contained in \eqref{fexp}. Evaluating
\eqref{P0exp} and \eqref{Ds0} for the special cases $\cF^{(0)}_s(p^2) \propto p^2$ and $F(p^2) \propto (p^2)^{N_{\rm max}}$, a simple
scaling argument establishes
\be\label{Dslimits}
\begin{split}
& \lim_{T \rightarrow \infty} \, D_S^{(0)}(T) = 4 \, , \qquad a_1 > 0 \, , \\
& \lim_{T \rightarrow 0} \, D_S^{(0)}(T) = \frac{4}{N_{\rm max}} \, .  
\end{split}
\ee
Thus $a_1 > 0$ ensures that the spectral dimension seen by particles for long
diffusion times coincides with the topological dimension of spacetime. Including
higher powers of momenta decreases $D_S^{(0)}(T)$ for short diffusion times. The
generalized spectral dimension then interpolates smoothly between these limits. This
feature is illustrated in Fig.\ \ref{figsd_F012_D}.
The case $N_{\rm max} = 1$, $a_1 > 0$ leads to a spectral dimension
which is independent of the diffusion time (dashed line). The spectral dimension
obtained for $N_{\rm max} = 3$ and momenta $Q_n = 1$, $n=1,0,-1,-2$. In this case,
$D_S^{(0)}(T)$ interpolates between $4$ at large $T$ and $4/3$ for small $T$ respectively. 
The crossover occurs for $T/ { t} \approx (4\pi)^2$ (Spin 0 and 1), resp., 
$T/ { t}^2 \approx 10 \times (4\pi)^2$ (Spin 2). Notably, it is only the shape 
of this crossover, which depends on the spin of the particle, while the asymptotic 
limits are universal for all spins. 

\subsection{Propagators with full momentum dependence}
\label{sect.IVb}
We now take the step beyond effective field theory and investigate
the spectral dimension arising from the inverse propagators \eqref{F2}
including the full momentum dependence. Comparing Figs.\ \ref{figsd_F012_F} and \ref{proptrunc}
the structural difference is immediate: in the effective field theory framework
$F_s(z)$ diverges as $z \rightarrow \infty$ while the inclusion
of the full momentum dependence renders $\lim_{z \rightarrow \infty} F_s(z)$ finite
with the limit given by the leading term in \eqref{latetime}. As a consequence of the modified
asymptotics, the integral \eqref{P0exp} diverges at the upper boundary, since the contribution
of large momenta is no longer exponentially suppressed once the full momentum dependent propagators 
are considered.

In order to still be able to analyse the spectral dimension arising
in this framework, we regulate  \eqref{P0exp} by introducing
a UV cutoff $\Lambda_{\rm UV}$,
\be\label{P0reg}
P^{(0)}(T; \Lambda_{\rm UV}) = \int^{\Lambda_{\rm UV}} \frac{d^4p}{(2\pi)^4} \, e^{-T F^{(0)}(p^2)} \, .
\ee
In the spirit of the discussion leading to eq.\ \eqref{d0def}, we consider
the ``massless'' structure functions $F_s^{(0)}(z)$
where the constant terms appearing in the late-time expansion \eqref{latetime}
have been removed. The return probability \eqref{P0reg} then allows to
construct the spectral dimension as a function of $\Lambda_{\rm UV}$
\be
D^{(0)}_S(T; \Lambda_{\rm UV}) = - 2 \, T \, \p_T \, \ln(P^{(0)}(T; \Lambda_{\rm UV})) \, . 
\ee
 The scale $\Lambda$ appearing in the spectral action \eqref{SpAct1} is thereby held fixed and sets the transition scale between the UV and IR regime. A detailed analytical and numerical analysis based on
the expansion \eqref{latetime} then establishes
\be
\begin{split}
\lim_{\Lambda_{\rm UV} \rightarrow \infty} D^{(0)}_S(T; \Lambda_{\rm UV}) = 
  \lim_{\Lambda_{\rm UV} \rightarrow \infty} \, 4 T \, F_s^{(0)}(\Lambda_{\rm UV}) \, . 
\end{split}
\ee
Based on the late-time expansion \eqref{latetime}, we thus we conclude that including the full momentum dependence
in the structure function leads to a spectral dimension which \emph{vanishes for all diffusion times $T$}:
\be
D^{(0)}_S(T) = \lim_{\Lambda_{\rm UV} \rightarrow \infty} D^{(0)}_S(T; \Lambda_{\rm UV}) = 0 \, .
\ee
This result entails in particular that there is no scaling regime for large diffusion time
where the spectral dimension matches the topological dimension. Thus including the full momentum dependence,
one does not recover a ``classical regime'' were the spectral dimension would indicate the onset
of classical low-energy physics.

\section{Conclusions and Outlook}
\label{sec:Concl}
In this work we constructed the generalized spectral dimension $D_S(T)$ describing the propagation 
of (massless) scalars, vectors and gravitons based on the \emph{classical} spectral action \eqref{SpAct1}. 
Our results distinguish three cases:
\begin{enumerate}
\item If the spectral action is interpreted as an effective field theory restricted to the 
power-counting relevant and marginal terms, the generalized spectral dimension is independent 
of the diffusion time $T$ and matches the topological dimension of spacetime $D_S(T) = 4$.
\item If the effective field theory framework is extended to also include power-counting irrelevant
terms, the generalized spectral action interpolates between $D_S(T) = 4$ for long diffusion time 
and $D_S(T) = 4/N_{\rm max}$ for short diffusion times. $N_{\rm max}$ is determined by the highest 
power of momentum contained in the propagator, $(p^2)^{N_{\rm max}}$. The crossover between these 
two asymptotic regimes is set by the cutoff $\Lambda$ and its shape explicitly depends on the spin 
of the propagating particle.
\item If the full momentum-dependence of the propagators is taken into account, the generalized 
spectral dimension becomes independent of the spin and vanishes identically $D_S(T) = 0$.
\end{enumerate}
The last feature can be traced back to the fact that the full propagators approach a constant for 
momenta much larger than the characteristic cutoff scale $\Lambda$. In Ref.\ \cite{Kurkov:2013kfa} 
the peculiar behaviour was summarized by the catchy phrase that ``high-energy bosons do not 
propagate''. Indeed the vanishing of the spectral dimension suggests that the momentum space of the 
theory resembles the one of a zero-dimensional field theory. In this sense the situation
resembles a picture where spacetime fragments into isolated points which do not communicate.

We stress, however, that all computations carried out in this work are at the classical level. 
In particular the spacetime is given by classical Euclidean space. All effects are due to the 
change of measure in momentum space reflected by non-canonical form of the classical propagators 
and thus do not capture properties of an underlying quantum spacetime. Nevertheless, we believe
that the spectral action, comprising the three scenarios discussed above, 
provides a valuable testbed for new candidates for quantum gravity observables generalizing the spectral dimension.

{

The vanishing of the spectral dimension, $D_S(T) = 0$, is in agreement with the previous computations
obtained for other non-commutative spacetimes \cite{Alesci:2011cg}.
This is a welcome and surprising result, since the non-commutative nature of our spectral triple construction
differs substantially from that of \cite{Alesci:2011cg}.
Aside for the limiting case of vanishing spectral dimension, both results display the same qualitative features:
The spectral dimension interpolates between the topological dimension and zero,
and has a local maximum situated close to a transition scale.
Additionally, the onset of the transition is, in both cases, controlled by the parameter
which is introduced by the non-commutative description
($\Lambda$ in our computation and the parameter $\kappa$
that governs the non-commutativity of the coordinates in \cite{Alesci:2011cg}).
Therefore, the transition is truly representative of a regime
in which the non-commutative features of the spacetime become important.

In agreement with the conjecture of ``Asymptotic Silence'' pushed forward in \cite{Carlip:2009kf},
it is a general feature of our computation that the spectral dimension decreases in the ultraviolet,
and thus describes a spacetimes that breaks into disconnected regions at high energies.
However, differently from \cite{Carlip:2009kf}, it is not immediately evident what
is the physical mechanism underlying this scenario.
Furthermore, our resuts are instead in stark contrast to 
the generalized spectral dimension typically obtained within quantum gravity approaches where $D_S(T)$ 
typically interpolates between $D_S(T) = 4$ (classical, macroscopic phase) and $D_S(T) < 4$ at 
microscopic scales \cite{Carlip:2009kf}.

At this stage, it is tempting to speculate 
that the vanishing spectral dimension is an artefact of extrapolating the classical spectral action 
into the trans-Planckian regime without taking quantization effects into account.
In other words, vacuum fluctuations seem to be fundamental to balance the fragmentation of spacetime
in the ultraviolet \cite{Carlip:2012md}.
An UV completion of the spectral action could be achieved through the Asymptotic Safety mechanism 
\cite{Codello:2008vh,Reuter:2012xf,Reuter:2012xf}, which seems a natural choice given the field 
content and symmetries of the model, may lead to $D_S(T)$ interpolating from four to two, 
while still keeping the non-polynomial momentum dependence in the propagators. We hope to come back to this intriguing 
possibility in the future.

}


\section*{Acknowledgments}
We thank M.A.\ Kurkov, Fedele Lizzi and Martin Reuter for inspiring discussions and Walter van Suijlekom for 
helpful comments and a critical reading of the
manuscript. N.A.\ and F.S.\ are financially supported by the Netherlands Organization for Scientific
Research (NWO) within the Foundation for Fundamental Research on Matter (FOM) grants
13PR3137 and 13VP12. O.Z. acknowledges financially supported by the DFG within the 
Emmy-Noether program (Grant SA/1975 1-1) while this work was in preparation. 


\providecommand{\href}[2]{#2}\begingroup\raggedright
\endgroup


\begin{thebibliography}{10}



\bibitem{Chamseddine:1991qh}
A.~H.~Chamseddine and A.~Connes,
Phys.\ Rev.\ Lett.\  {\bf 77} (1996) 4868.
  
\bibitem{Chamseddine:1996zu}
A.~H.~Chamseddine and A.~Connes,
Commun.\ Math.\ Phys.\  {\bf 186} (1997) 731
[hep-th/9606001].


\bibitem{Connes:1994yd} 
A.~Connes,
``Noncommutative geometry,'' Academic Press, San Diego (1994). 

\bibitem{connesreview}
A.~Connes, {\it Noncommutative geometry and physics}, in
B.~Julia, J.~Zinn-Justin (Eds.),
Proc. 1992 Les Houches Summer School, North-Holland, Amsterdam (1995).

\bibitem{Jureit:2007qm}
  J.~H.~Jureit, T.~Krajewski, T.~Schucker and C.~A.~Stephan,
  Acta Phys.\ Polon.\ B {\bf 38} (2007) 3181
  [arXiv:0705.0489 [hep-th]].
    
\bibitem{Sakellariadou:2013ve}
  M.~Sakellariadou,
  J.\ Phys.\ Conf.\ Ser.\  {\bf 442} (2013) 012015
  [arXiv:1301.4687 [hep-th]].

\bibitem{vandenDungen:2012ky} 
K.~van den Dungen and W.~D.~van Suijlekom,
Rev.\ Math.\ Phys.\  {\bf 24} (2012) 1230004 
[arXiv:1204.0328 [hep-th]].


\bibitem{Connes:2006qv}
  A.~Connes,
  JHEP {\bf 0611} (2006) 081
  [hep-th/0608226].

\bibitem{Chamseddine:2006ep}
  A.~H.~Chamseddine, A.~Connes and M.~Marcolli,
  Adv.\ Theor.\ Math.\ Phys.\  {\bf 11} (2007) 991
  [hep-th/0610241].
  
\bibitem{Chamseddine:2007ia}
  A.~H.~Chamseddine and A.~Connes,
  Phys.\ Rev.\ Lett.\  {\bf 99} (2007) 191601
  [arXiv:0706.3690 [hep-th]].

  
  
\bibitem{Chamseddine:2012sw}
  A.~H.~Chamseddine and A.~Connes,
  JHEP {\bf 1209} (2012) 104
  [arXiv:1208.1030 [hep-ph]].
  
\bibitem{Stephan:2013rna}
  C.~A.~Stephan,
  arXiv:1305.3066 [hep-ph].
  

\bibitem{Chamseddine:2010ud}
  A.~H.~Chamseddine and A.~Connes,
  Fortsch.\ Phys.\  {\bf 58} (2010) 553
  [arXiv:1004.0464 [hep-th]].

\bibitem{Devastato:2013oqa}
  A.~Devastato, F.~Lizzi and P.~Martinetti,
  JHEP {\bf 1401} (2014) 042
  [arXiv:1304.0415 [hep-th]].

\bibitem{Chamseddine:2013sia}
  A.~H.~Chamseddine, A.~Connes and W.~D.~van Suijlekom,
  J.\ Geom.\ Phys.\  {\bf 73} (2013) 222
  [arXiv:1304.7583 [math-ph]].

\bibitem{Chamseddine:2013rta}
A.~H.~Chamseddine, A.~Connes and W.~D.~van Suijlekom,
JHEP {\bf 1311} (2013) 132
[arXiv:1304.8050 [hep-th]].


\bibitem{Ishihara:2013asa}
  S.~Ishihara, H.~Kataoka, A.~Matsukawa, H.~Sato and M.~Shimojo,
  arXiv:1311.4944 [hep-th].

\bibitem{Beenakker:2014yla}
  W.~Beenakker, W.~D.~van Suijlekom and T.~v.~d.~Broek,
  arXiv:1409.5982 [hep-th].
  
\bibitem{Beenakker:2014zla}
  W.~Beenakker, W.~D.~van Suijlekom and T.~v.~d.~Broek,
  arXiv:1409.5983 [hep-th].
  
\bibitem{Beenakker:2014ama}
  W.~Beenakker, W.~D.~van Suijlekom and T.~v.~d.~Broek,
  arXiv:1409.5984 [hep-th].


\bibitem{vanSuijlekom:2011uu}
 W.~D.~van Suijlekom,
 JHEP {\bf 1103} (2011) 146
 [arXiv:1101.4804 [math-ph]].

\bibitem{vanSuijlekom:2011kc}
 W.~D.~van Suijlekom,
 Commun.\ Math.\ Phys.\  {\bf 312} (2012) 883
 [arXiv:1104.5199 [math-ph]].

\bibitem{vanSuijlekom:2012xb}
 W.~D.~van Suijlekom,
 Phys.\ Lett.\ B {\bf 711} (2012) 434
 [arXiv:1204.4070 [hep-th]].

\bibitem{Estrada:2012te}
  C.~Estrada and M.~Marcolli,
  Int.\ J.\ Geom.\ Meth.\ Mod.\ Phys.\  {\bf 10} (2013) 1350036
  [arXiv:1208.5023 [hep-th]].

\bibitem{Suijlekom:2014ata} 
W.~D.~van Suijlekom,
Annales Henri Poincare {\bf 15} (2014) 985.


\bibitem{Nelson:2010ru}
  W.~Nelson, J.~Ochoa and M.~Sakellariadou,
  Phys.\ Rev.\ Lett.\  {\bf 105} (2010) 101602
  [arXiv:1005.4279 [hep-th]].

\bibitem{Lambiase:2013dai}
  G.~Lambiase, M.~Sakellariadou and A.~Stabile,
  JCAP {\bf 1312} (2013) 020
  [arXiv:1302.2336 [gr-qc]].

\bibitem{Chamseddine:2014nxa} 
A.~H.~Chamseddine, A.~Connes and V.~Mukhanov,
arXiv:1409.2471 [hep-th].




\bibitem{D'Andrea:2013nda}
  F.~D'Andrea, F.~Lizzi and P.~Martinetti,
  J.\ Geom.\ Phys.\  {\bf 82} (2014) 18
  [arXiv:1305.2605 [math-ph]].
  
\bibitem{Farnsworth:2013nza}
  S.~Farnsworth and L.~Boyle,
  arXiv:1303.1782 [hep-th].
  


       
\bibitem{Kurkov:2013kfa}
M.~A.~Kurkov, F.~Lizzi and D.~Vassilevich,
Phys.\ Lett.\ B {\bf 731} (2014) 311
[arXiv:1312.2235 [hep-th]].
  

\bibitem{Iochum:2011yq}
 B.~Iochum, C.~Levy and D.~Vassilevich,
 Commun.\ Math.\ Phys.\  {\bf 316} (2012) 595
 [arXiv:1108.3749 [hep-th]].
  
  


  
  
  
\bibitem{Reuter:2011ah}
M.~Reuter and F.~Saueressig,
JHEP {\bf 1112} (2011) 012
[arXiv:1110.5224 [hep-th]].

\bibitem{Ambjorn:2005db} 
J.~Ambj{\o}rn, J.~Jurkiewicz and R.~Loll,
Phys.\ Rev.\ Lett.\  {\bf 95} (2005) 171301
[hep-th/0505113].
    
\bibitem{Sotiriou:2011mu} 
T.~P.~Sotiriou, M.~Visser and S.~Weinfurtner,
Phys.\ Rev.\ Lett.\  {\bf 107} (2011)  131303 
[arXiv:1105.5646 [gr-qc]].

\bibitem{Carlip:2009kf} 
S.~Carlip,
arXiv:0909.3329 [gr-qc].
    
\bibitem{Carlip:2012md} 
S.~Carlip,
AIP Conf.\ Proc.\  {\bf 1483} (2012) 63 
[arXiv:1207.4503 [gr-qc]].


\bibitem{Benedetti:2009ge}
D.~Benedetti and J.~Henson,
Phys.\ Rev.\ D {\bf 80} (2009) 124036
[arXiv:0911.0401 [hep-th]].


\bibitem{Horava:2009if} 
P.~Horava,
Phys.\ Rev.\ Lett.\  {\bf 102}  (2009) 161301
[arXiv:0902.3657 [hep-th]].
\bibitem{Sotiriou:2011aa}
  T.~P.~Sotiriou, M.~Visser and S.~Weinfurtner,
  Phys.\ Rev.\ D {\bf 84} (2011) 104018
  [arXiv:1105.6098 [hep-th]].
\bibitem{Amelino-Camelia:2013gna} 
  G.~Amelino-Camelia, M.~Arzano, G.~Gubitosi and J.~Magueijo,
  Phys.\ Rev.\ D {\bf 88}  (2013) 103524
  [arXiv:1309.3999 [gr-qc]].

\bibitem{Amelino-Camelia:2013cfa} 
  G.~Amelino-Camelia, M.~Arzano, G.~Gubitosi and J.~Magueijo,
  Phys.\ Lett.\ B {\bf 736}  (2014) 317
  [arXiv:1311.3135 [gr-qc]].

\bibitem{Calcagni:2010pa} 
G.~Calcagni,
Phys.\ Lett.\ B {\bf 697}  (2011) 251
[arXiv:1012.1244 [hep-th]].

\bibitem{Calcagni:2013sca} 
G.~Calcagni and G.~Nardelli,
Phys.\ Rev.\ D {\bf 88} (2013)  124025 
[arXiv:1304.2709 [math-ph]].
  
\bibitem{Calcagni:2013vsa}
G.~Calcagni, A.~Eichhorn and F.~Saueressig,
Phys.\ Rev.\ D {\bf 87} (2013) 12,  124028
[arXiv:1304.7247 [hep-th]].


\bibitem{Lauscher:2005qz} 
O.~Lauscher and M.~Reuter,
JHEP {\bf 0510}  (2005) 050
[hep-th/0508202].

\bibitem{Reuter:2012xf} 
M.~Reuter and F.~Saueressig,
Lect.\ Notes Phys.\  {\bf 863}  (2013) 185
[arXiv:1205.5431 [hep-th]].

  
\bibitem{Rechenberger:2012pm} 
  S.~Rechenberger and F.~Saueressig,
  Phys.\ Rev.\ D {\bf 86}  (2012) 024018
  [arXiv:1206.0657 [hep-th]].

\bibitem{Modesto:2008jz} 
L.~Modesto,
Class.\ Quant.\ Grav.\  {\bf 26}  (2009) 242002
[arXiv:0812.2214 [gr-qc]].

\bibitem{Modesto:2009qc} 
L.~Modesto and P.~Nicolini,
Phys.\ Rev.\ D {\bf 81}  (2010) 104040
[arXiv:0912.0220 [hep-th]].

\bibitem{Magliaro:2009if} 
E.~Magliaro, C.~Perini and L.~Modesto, \\
arXiv:0911.0437 [gr-qc].

\bibitem{Caravelli:2009gk} 
F.~Caravelli and L.~Modesto,
arXiv:0905.2170 [gr-qc].
  
\bibitem{Modesto:2009kq} 
L.~Modesto,
arXiv:0905.1665 [gr-qc].

\bibitem{Calcagni:2013dna} 
G.~Calcagni, D.~Oriti and J.~Th{\"u}rigen,
Class.\ Quant.\ Grav.\  {\bf 31} (2014) 135014
[arXiv:1311.3340 [hep-th]].

\bibitem{Eichhorn:2013ova} 
  A.~Eichhorn and S.~Mizera,
  Class.\ Quant.\ Grav.\  {\bf 31} (2014) 125007 
  [arXiv:1311.2530 [gr-qc]].

\bibitem{Benedetti:2008gu} 
D.~Benedetti,
Phys.\ Rev.\ Lett.\  {\bf 102} (2009) 111303 
[arXiv:0811.1396 [hep-th]].

\bibitem{Arzano:2014jfa}
  M.~Arzano and T.~Trzesniewski,
  Phys.\ Rev.\ D {\bf 89} (2014) 124024
  [arXiv:1404.4762 [hep-th]].

\bibitem{Alesci:2011cg} 
 { 
  E.~Alesci and M.~Arzano,
  Phys.\ Lett.\ B {\bf 707}, 272 (2012)
  [arXiv:1108.1507 [gr-qc]].
  }

\bibitem{Codello:2008vh}
A.~Codello, R.~Percacci and C.~Rahmede,
Annals Phys.\  {\bf 324} (2009) 414
[arXiv:0805.2909 [hep-th]].
    
\bibitem{Barvinsky:1987uw} 
A.~O.~Barvinsky and G.~A.~Vilkovisky,
Nucl.\ Phys.\ B {\bf 282} (1987) 163.

\bibitem{Barvinsky:1990up} 
A.~O.~Barvinsky and G.~A.~Vilkovisky,
Nucl.\ Phys.\ B {\bf 333} (1990) 471.

\bibitem{Codello:2012kq} 
A.~Codello and O.~Zanusso,
J.\ Math.\ Phys.\  {\bf 54} (2013)  013513
[arXiv:1203.2034 [math-ph]].

\bibitem{Vassilevich:2003xt}
  D.~V.~Vassilevich,
  Phys.\ Rept.\  {\bf 388} (2003) 279
  [hep-th/0306138].

\bibitem{Devastato:2013wza}
  A.~Devastato,
  Phys.\ Lett.\ B {\bf 730} (2014) 36
  [arXiv:1309.5973 [hep-th]].

\bibitem{Calcagni:2014wba}
G.~Calcagni, L.~Modesto and G.~Nardelli,
arXiv:1408.0199 [hep-th].

\bibitem{Carlip:2011tt} 
 { 
  S.~Carlip, R.~A.~Mosna and J.~P.~M.~Pitelli,
  Phys.\ Rev.\ Lett.\  {\bf 107}, 021303 (2011)
  [arXiv:1103.5993 [gr-qc]].
  }









\end{thebibliography}
\end{document}